\documentclass{mn2e}

\def\ltsim{\lower.5ex\hbox{$\; \buildrel < \over \sim \;$}}
\def\gtsim{\lower.5ex\hbox{$\; \buildrel > \over \sim \;$}}
\def\ltsim{\lower.5ex\hbox{$\; \buildrel < \over \sim \;$}}
\def\gtsim{\lower.5ex\hbox{$\; \buildrel > \over \sim \;$}}

\def\vx{{\bf x}}
\def\vq{{\bf q}}
\def\vv{{\bf v}}
\def\dd{{\rm d}}
\def\hmpc{{ {\rm h}^{-1} {\rm Mpc} }}
\def\hmpcs{{ {\rm h}^{-2} {\rm Mpc^2} }}

\usepackage{epsfig}
\newcommand{\vtheta}{\mbox{\boldmath{$\theta$}}}
\newcommand{\etal}{et al. }
\newcommand{\mnras}{MNRAS}
\newcommand{\apj}{ApJ}

\title{Halo Assembly Bias in the Quasi-linear Regime}

\author[J.~A. Keselman \& A. Nusser]  {Jose Ariel Keselman\thanks{Email:
        kari@tx.technion.ac.il} and Adi Nusser \\
        Physics department, IIT, Technion city, Haifa,
        Israel}

\begin{document} \maketitle

\begin{abstract}
We address the question of whether or not assembly bias
arises in the absence of highly non-linear effects such as tidal stripping
of halos near larger mass concentrations.
Therefore, we use a simplified dynamical scheme where these  effects 
are not modeled. We choose the  
punctuated Zel'dovich (PZ) approximation, which prevents orbit mixing  by coalescing particles coming within a critical distance  of each other. 
A numerical implementation of this approximation is fast, allowing us to run a 
large number of simulations to study assembly bias. 
We measure an assembly bias  from 60 PZ  simulations, each  with $512^3$
cold particles in a $128h^{-1}{\rm Mpc}$ cubic box. 
The assembly bias
estimated from the correlation functions at separations of $\ltsim 5h^{-1}{\rm Mpc}$
for objects (halos) at  $z=0$  is comparable to that  obtained in  full N-body simulations. 
For masses  $4\times 10^{11}h^{-1}M_\odot$ the ``oldest'' $10\%$ haloes are 
3-5 times more correlated than the ``youngest'' $10\%$. The bias weakens
with increasing mass, also in agreement with  full N-body simulations. 
We find that halo ages 
are correlated with the dimensionality of the surrounding linear structures 
as measured by the parameter $(\lambda_1+\lambda_2+\lambda_3)/
(\lambda_1^2+\lambda_2^2+\lambda_3^2)^{1/2}$ where $\lambda_i$ are proportional to the eigenvalues 
of the velocity deformation tensor. 
Our results suggest
that assembly bias may already be encoded in the early stages of the evolution.
\end{abstract}


\section{INTRODUCTION}
A fundamental question in cosmology is the relation between the galaxy
distribution and the underlying density field of the gravitationally dominant
dark matter. According to the standard paradigm of structure formation, galaxies
are harbored in stable virialized objects (halos) made of dark matter particles. An assumption that has been often made is that the clustering properties of halos depend on
their mass alone.
Although the   assumption seems over-simplistic given the complexity of 
the hierarchical process of halo formation, it is   sustained by the excursion set theory
(Bond \etal 1991; Lacey \& Cole 1993; Mo \& White 1996) and by results of
N-simulations of intermediate resolution  (Lemson \& Kauffmann 1999; 
Percival \etal 2003). Only recently  Gao, Springel \& White (2005) 
used a  simulation of 
exceptionally large dynamical range (the Millennium Simulation, Springel et al. 2005)
to show that the clustering of halos depend also on the their age, which is defined as the time
since a halo acquired half of its current mass. They have found, that the  ``oldest'' $10\%$
of the halos with mass $10^{11}h^{-1}M_{\odot}$ are more than 5 times more correlated then
the ``youngest'' $10\%$ halos of the same mass. This assembly bias has been
confirmed by Harker \etal (2006) using marked correlation functions on the same
simulation, and by Wechsler \etal (2006) and Jing \& Mo (2006) using independent
simulations. Wetzel \etal (2007) also found dependence of clustering on halo history,
but only when using a different definition for the assembly redshift.

\begin{figure*}
\centerline{\epsfig{figure=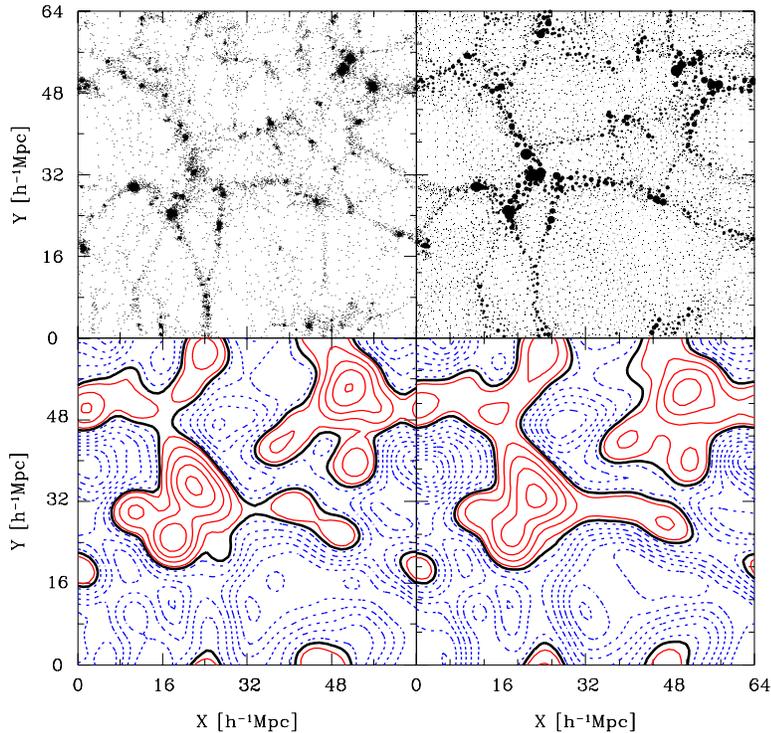, width=300pt, height=295pt}}
\caption{A visual comparison between results of the  PZ approximation (right)
and a PM N-body  code (left) run with the same initial conditions for $256^3$
particles in a box of $64{\rm Mpc h}^{-1}$ on the side. {\it Top: } particle
(object for PZ) distributions  in a slice of $3.2{ \rm Mpc}h^{-1}$ in thickness
(for the PM only a random subset of all the particles is shown). For the PZ,
each object is represented as a filled circle with radius proportional to the
mass. {\it Bottom:} contours maps of $\log_{10}\rm  (density)$ in the same slice.
The density fields are smoothed with a Gaussian window of a width of
$1.125{ \rm Mpc}h^{-1}$. The thin solid and dashed lines are density contours
above and below the mean, respectively.  Thick solid lines indicate mean density
contours. The contour spacing is 0.18. }
\label{fig:vis}
\end{figure*}

We still lack a completely satisfactory explanation for the origin of assembly bias. For gaussian initial conditions, simple arguments based on the spherical collapse model applied 
to narrow and broad initial density peaks which would collapse to halos of the same mass at the present time do predict an assembly bias, but 
with younger halos being more clustered than older ones. This trend 
of the bias is opposite to  what is seen in simulations. Tidal stripping has been
 also invoked as a possible mechanism (e.g. Diemand, Kuhlen \& Madau 2007). Because of mass
stripping by the tidal gravitational field of the large mass concentration, nearby
halos would have been of higher mass in a different environment. Therefore, these halos would
 have earlier formation times and would be more biased than halos of the
same mass in the field. Avila-Reese \etal (2005) suggested tidal stripping as a mechanism responsible for generation of assembly bias in the high density regions whereas in low-density regions the cosmological initial conditions play a more important role. Maulbetsch \etal (2007) and Wang, Mo \& Jing (2006) and Desjacques (2007)
suggest that the halo mass-accretion is less efficient in denser regions due to large scale tidal fields.  However, the extent of this effect is difficult to assess.
Here we examine whether the  bias can, at least partially, arise in 
the quasi-linear evolution (i.e. over scales where the flow 
is still laminar).
In order to eliminate highly non-linear effects such as tidal stripping, we adopt approximate methods 
based on the Zel'dovich approximation (Zel'dovich 1970) where 
particles move on
straight lines independent of the motion of other particles.  The Zel'dovich
approximation is an analytic solution to planar cosmological perturbations up to
the stage where multi-streaming appears. For three dimensional perturbations,
the approximation is a reasonable description of quasi-linear dynamics 
away from multi-streaming regions (e.g. Nusser \etal 1991).
In order to extend the applicability of this approximation beyond multi-streaming,
we adopt the following scheme. Particles initially move in straight lines according
to Zel'dovich, but they are merged together when they  come within a critical
distance of each other. This merging (sticking) produces an object with mass
and linear momentum equal the total of its components. The critical distance is
taken to depend on time like a diffusion length, as inspired by the adhesion
approximation.  This is known as the punctuated Zel'dovich (PZ) approximation (Fontana \etal 1995).
This approximation is ideal for our purposes as it 
does not incorporate highly nonlinear effects and 
also it is fast and easy to implement. Further, it readily provides merging trees for
individual objects. 

The outline of the paper is as follows. In \S\ref{sec:app} we discuss 
the Zel'dovich approximation and our implementation of 
the punctuated Zel'dovich (PZ) scheme to describe the evolution in the quasi-linear regime. 
In \S\ref{sec:simu} we present the PZ simulations and compare them  with 
results from full dynamics.
Our  main results for  halo biasing in  the PZ simulations are presented in 
\S\ref{sec:results}. We conclude with a general discussion in  \S\ref{sec:sum}

\section{The Approximate Dynamics}
\label{sec:app}
According to the Zel'dovich approximation,
the time dependent  position, $\vx$,  of a particle with initial (Lagrangian)
coordinate $\vq$ is
\begin{equation}
\vx=\vq+D(t)\vtheta(\vq) \; ,
\label{eq:za}  
\end{equation}
where we work with comoving coordinates, the function $D(t)$ is the gravitational growth rate of linear density fluctuations,
and $\vtheta(q)$ is a vector field which is a function of $\vq$ only and is
assumed to be derived from a potential. The physical peculiar velocity
is $\vv=a(t)\dd \vx /\dd t=a \dot D\vtheta$, where $a(t)$ is the expansion
factor of the cosmological background. The relation  (\ref{eq:za}) yields a reasonable 
description of the gravitational evolution in the quasi-linear regime  (e.g. Nusser \etal 1991; Weinberg \& Gunn 1990), but fails 
near collapsed regions where multi-streaming occurs.
The punctuated Zel'dovich (PZ) approximation is  an
extension of (\ref{eq:za}), which prevents the coasting away of particles in 
collapsed regions and yet preserves the simplicity of the Zel'dovich kinematics. In the PZ,
one starts with equal mass particles located at a uniform cubic grid at some
initial time $t_{\rm i}\rightarrow 0$. A particle is displaced from an initial
position, $\vq$, at $t_{\rm i}$, according to (\ref{eq:za}) until it comes
within a distance of $d_{\rm c}$ of another particle. The two particles are
then merged  together to form an ``object" having their total mass and
linear momentum, and placed at their center of mass. 
The newly formed object 
 is then  moved according to the Zel'dovich relation (\ref{eq:za})
with $\vtheta $ determined from momentum conservation.  
The scheme  is applied to describe the further merging and evolution 
of objects (and particles).  

We  interpret the PZ  in the framework of the adhesion approximation
according to which (e.g. Shandarin 1991, Nusser \& Dekel 1990)
\begin{equation}
\frac{\dd \vtheta}{\dd D}=\nu \Delta \vtheta \; .
\label{eq:adhesion}
\end{equation}
The viscous term modifies the Zel'dovich ansatz $\vtheta=constant$ in regions
with high velocity gradients and prevents orbit crossing. Viscosity affects the
flow over (comoving) scales $\ltsim \sqrt{D \nu}$. Above these scales the flow is described
by the usual Zel'dovich approximation.
For  $d_{\rm c}\propto \sqrt{\nu D(t)}$, the PZ is reminiscent of the 
adhesion approximation  
except that  it  ignores the details of the flow on scale $\ltsim \sqrt{D \nu}$ by coalescing objects 
which  are within a distance $d_{\rm c}$ of each other. 
Here we work with $d_{\rm c}=\sqrt{\nu D}$, as motivated by the adhesion approximation.

\section{The simulations}
\label{sec:simu}

We have run  60 PZ simulations, each with $512^3$ equal mass
particles in a $128h^{-1}{\rm Mpc}$  cubic box on the side. 
The initial conditions
correspond to a random Gaussian realization of the Cold Dark Matter scenario
in a flat universe without a cosmological constant. 
The dependence on the background density parameters comes through the 
initial power spectrum but 
the  dynamics is nearly independent of these  parameters when the linear growth factor is used as the time variable (Nusser \& Colberg 1998).
Therefore, apart from the effect of the initial power spectrum, the final result 
should be independent of the cosmological background.
Thus the particle mass is $4.3\times 10^9h^{-1}M_{\odot}$. 
The dimensionless value of the Hubble constant is $h=0.73$ and the {\it rms} 
value of the initial mass fluctuation in a sphere of $8\hmpc$, 
as extrapolated to current time, is $\sigma_8=0.8$.
The particles are moved forward in time from $z=1000$ to $z=0$ according to
the PZ approximation with $d_{\rm c}=\sqrt{\nu D}$ with $\nu=1\hmpcs$
which sets a spatial resolution of $1\hmpc$ at the present time ($D=1$).
To
further improve performance of the PZ, we smooth the initial velocity and density fields with a Gaussian window of width equal to  $\sigma=1h^{-1}{\rm Mpc}$.

Our time variable is the growth factor $D$ and the 
time-step, $\dd D$,  is such that $|\theta_{\rm max}| \dd D=0.1 d_{\rm c}$,
where $\theta_{\rm max}$ is the speed of the fastest object.

For purposes of history tracking, each object is assigned a unique
ID. When objects merge, the newly formed objects 
inherits the ID of the most massive progenitor. 
Object histories are tabulated in time slices separated by $\delta D=1/100.$

At the final time,
 the average number of objects per simulation is  $~7\times 10^5$ 
with  $~10^5$ being more massive than $4.3 \times 10^{11}{\rm h}^{-1} M_\odot$ which is the minimal halo mass we consider for the study of assembly bias.
The number of halos in all 60 simulations is comparable to that in 
the millennium simulation. 

\begin{figure}
\centerline{\epsfig{figure=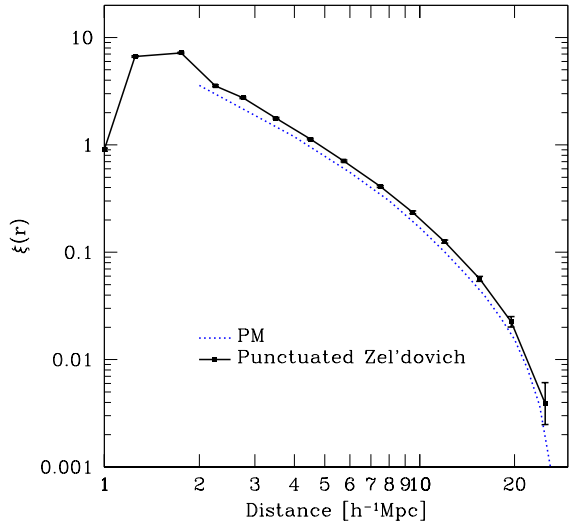, width=226pt, height=214pt}}
\caption{
Density correlation functions computed from the PZ and PM simulations, as
indicated in the figure. The mean of correlation functions from 60 PZ runs
(each of $512^3$ particles in a box of 128 $\hmpc$ on the side)  is shown as
the solid line. The attached error-bars are $1\sigma$. For this comparison we
use  a PM simulation of $256^3$ particles in a box size of $512 \hmpc$ on the
side. }
\label{fig:corr_func}
\end{figure}

The PZ approximation is neither expected nor intended  to
model highly non-linear dynamics. Indeed, it is because 
it misses highly non-linear effects that we use it in this study.
However,  it is prudent to make  a general  comparison  of our implementation of the PZ scheme  with results from full dynamics. 
To make a direct comparison, we have run the PZ scheme and a Particle-Mesh (Bertschinger \& Gelb 1991) 
N-body code on the same initial conditions for $256^3$ particles in cubic box of $64\hmpc$ on the side.
The initial conditions correspond to a flat CDM universe without a cosmological constant and $\sigma_8=0.8$.
Figure (\ref{fig:vis})
offers a visual impression of the difference between the final 
results from the PZ (panels to the right) and PM simulations (to the left). 
 In the top panels
the final distribution of objects  in the PZ
approximation is seen to follow  closely the particle distribution in the PM code. This
impression is further confirmed by the contour maps of the density fields 
 in the bottom panels. The general agreement is impressive. The density fluctuations in the PZ are slightly of
larger amplitude but this maybe due to cosmic variance.
 In figure
(\ref{fig:corr_func}) we also plot the density correlation functions obtained
from PZ runs and a  PM simulation. The solid line is the mean correlation
computed from the mean of 60 PZ runs (each of $512^3$ particles in a box of
$128\hmpc$ on the side), while the dashed is computed from the output of a
single PM simulation of $256^3$ particles in a box of $512\hmpc$ on the side.
The $1\sigma$ error-bars attached to the PZ curve are estimated using the 
bootstrap method as follows. 
We generate 500 sets of simulations where each set contains  
60 simulations picked  randomly out of the 60 original simulations (i.e. some of 
these simulations could be  selected more than once).
For each set we compute the mean  correlation  and the errors are estimated as
the standard deviations between the mean correlations of the 500 sets.  

\begin{figure*}
\centerline{\epsfig{figure=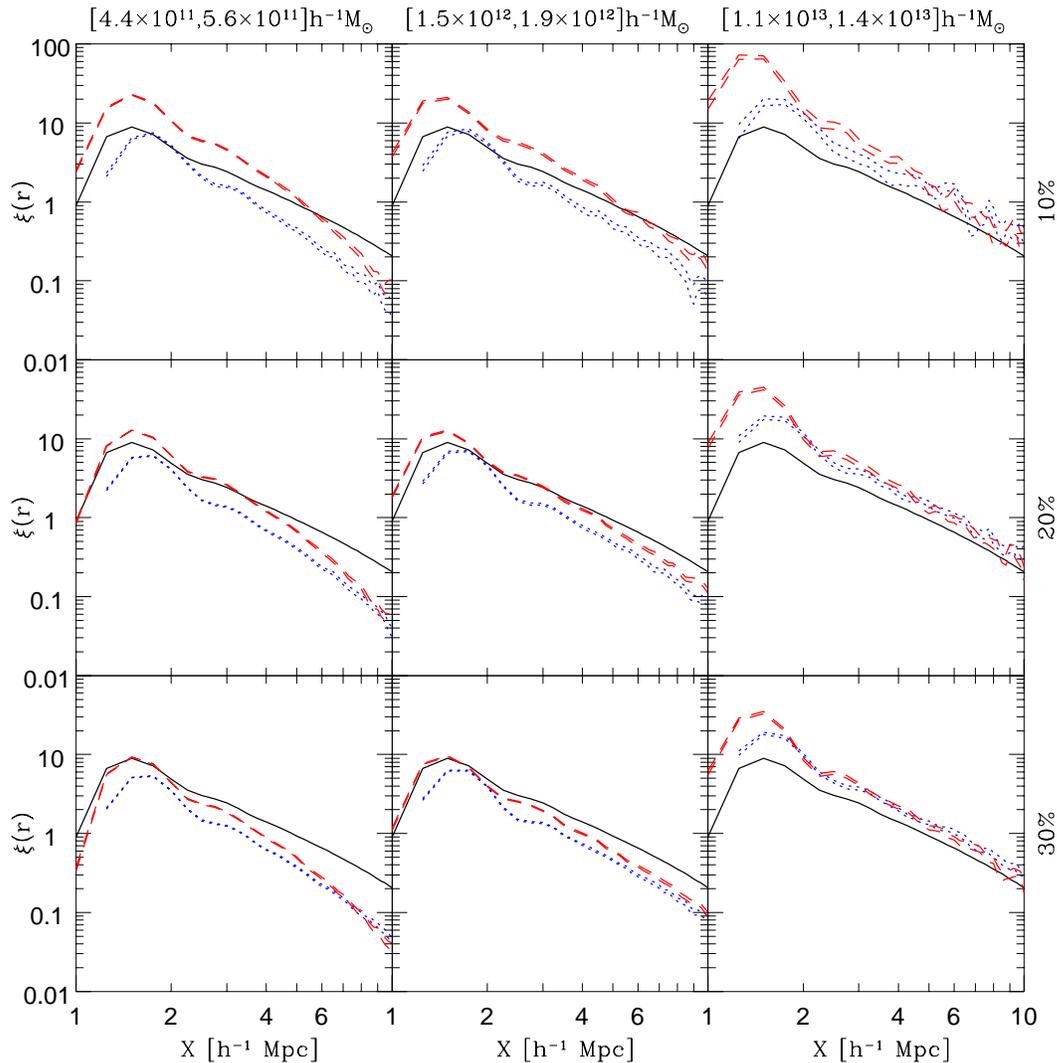, width=400pt, height=400pt}}
\caption{Correlation functions  for $10\%$, $20\%$ and
$30\%$ oldest/youngest haloes. In each panel, the correlations are 
represented by curves corresponding to   $\pm\sigma$ deviations.
Dashed and dotted lines are  for old and young halos, respectively. 
For comparison, the solid line in each panel 
shows the correlation function of the underlying mass- density.
}
\label{fig:sets_corr_func}
\end{figure*}

The bump in $\xi(r)$ in figure (\ref{fig:corr_func}) at scales smaller than $2\hmpc$
(also visible in figure \ref{fig:sets_corr_func}) is due to the finite resolution  in the simulations. Therefore, we will base our conclusion on 
correlations on  scales larger than $2\hmpc$

Objects (``halos'') in a PZ simulation are point-like and are identified 
using different criteria than halos in full N-body simulations. 
Therefore, we expect only a rough agreement between the  mass functions 
of halos (number density versus mass)  computed from  PZ simulations and 
  full dynamics. 
We compared the abundance of objects versus mass in the PZ runs with the analytic predictions of Sheth \& Tormen (2002) and Press \& Schechter (1974) for the halo mass function. The transfer function used in these predictions
is taken from Bardeen \etal (1986) with a slope of $n=1$ for  the primordial power spectrum.

Overall, there is only a qualitative agreement between PZ and the analytic expressions.    For masses $~4.3 \times 10^{11}h^{-1}M_\odot$ the PZ simulation agrees with PS and ST. However, for more massive haloes, the PZ overestimates abundance up to a factor of two for  masses $~10^{13}h^{-1}M_\odot$. The difference is reduced 
as we go to  higher masses until it disappears at  $6 \times 10^{13}h^{-1}M_\odot$.
 At higher masses, PZ falls short of the analytic expressions by 
 a factor which increases with mass.

\section{Results}

\label{sec:results}

\begin{figure*}
\centerline{\epsfig{figure=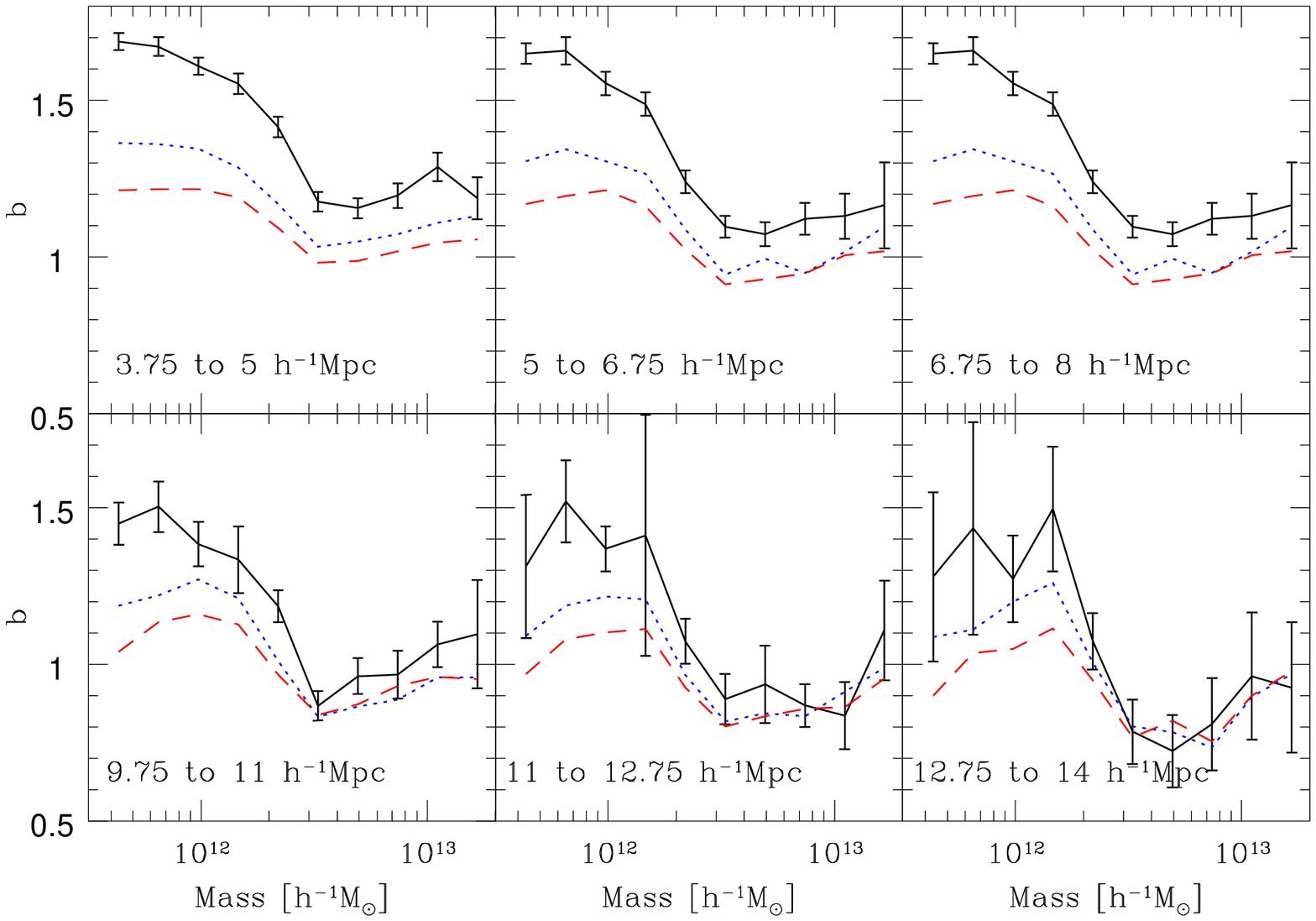, width=400pt, height=284pt}}
\caption{Square root ratio of clustering amplitude of  old  to young halos as a function of their mass
at various separations, as indicated in the panels. 
Solid, dotted and dashed curves, respectively,
 correspond to $10\%$, $20\%$ and $30\%$ oldest/youngest haloes.
Error-bars ($1\sigma$) are plotted only for the solid lines.
The value of $M_\star$ corresponding to our initial conditions is $5\times 10^{12}h^{-1} M_\odot$.}
\label{fig:sets_bias}
\end{figure*}

The merging history of an object (``halo'') in our implementation of the PZ approximation 
is readily provided.   We consider only halos containing more than 100 particles ($4.3\times 10^{11}h^{-1}M_\odot$) at the final time $(z=0)$ and
define the formation time of a halo as the redshift at which 
it has acquired half of its final mass (Gao, Springel \& White  2005).
We use the correlation functions to probe the clustering properties of 
halos.
In figure \ref{fig:sets_corr_func} we plot the correlation functions, $\xi(x)$, 
as a function of separation, $x$,  for 
halos in three mass ranges in the left, middle and right columns, respectively.
The dashed (dotted) lines in the  top, middle and bottom panels, 
respectively,  
correspond to 10\%, 20\% and 30\%  oldest (youngest) halos. 
The solid lines in all panels are identical and represent the 
correlation  function of the mass density field. In each panel the halo correlations are 
shown by two curves representing $\pm\sigma$ deviations computed using the bootstrap method, as outlined in \S\ref{sec:simu}.
The dependence of the correlation function on the formation time
is clear for all mass ranges shown in the figure.
The bias persists even between the 30\% youngest and 30\% oldest halos. 
  
We use the difference between the correlation functions of old and young halos to 
quantify the assembly bias at various separations. 
We determine   the bias parameter $b$ in separation range $(x,x+\Delta x)$ by 
minimizing  the quantity (Gao \& White 2007)
\[ \int_{\Delta x} \dd x \left[ \log~\xi_{\rm old}-\log(b^2\times \xi_{\rm young})\right]^2\; .\]
Figure (\ref{fig:sets_bias}) shows the bias as a function of
 halo mass, for various separations.
 For masses $\ltsim 2-3\times 10^{12}M_\odot$, the bias 
 is about $1.7$ and is similar for all separations considered here. 
 The error-bars are large at separations $>10\hmpc$ and 
 we cannot detect an increase in the bias as claimed by Gao, Springel \& White (2005).
 The bias weakens with increasing halo mass, but remains statistically significant 
 only for the 10\% old/young halos, at separations $\ltsim 8\hmpc$.
 The figure shows that the mass scale $2-3\times 10^{12}M_\odot$ marks 
 a mass threshold above which assembly bias weakens, for all separations. 
 This threshold  is close  to the non-linear mass scale
 $M_\star$ defined as the mass scale over which the rms of density fluctuations 
 is $1.69$. For our initial conditions $M_\star \approx 5\times 10^{12}h^{-1}M_\odot$.

\begin{figure*}
\centerline{\epsfig{figure=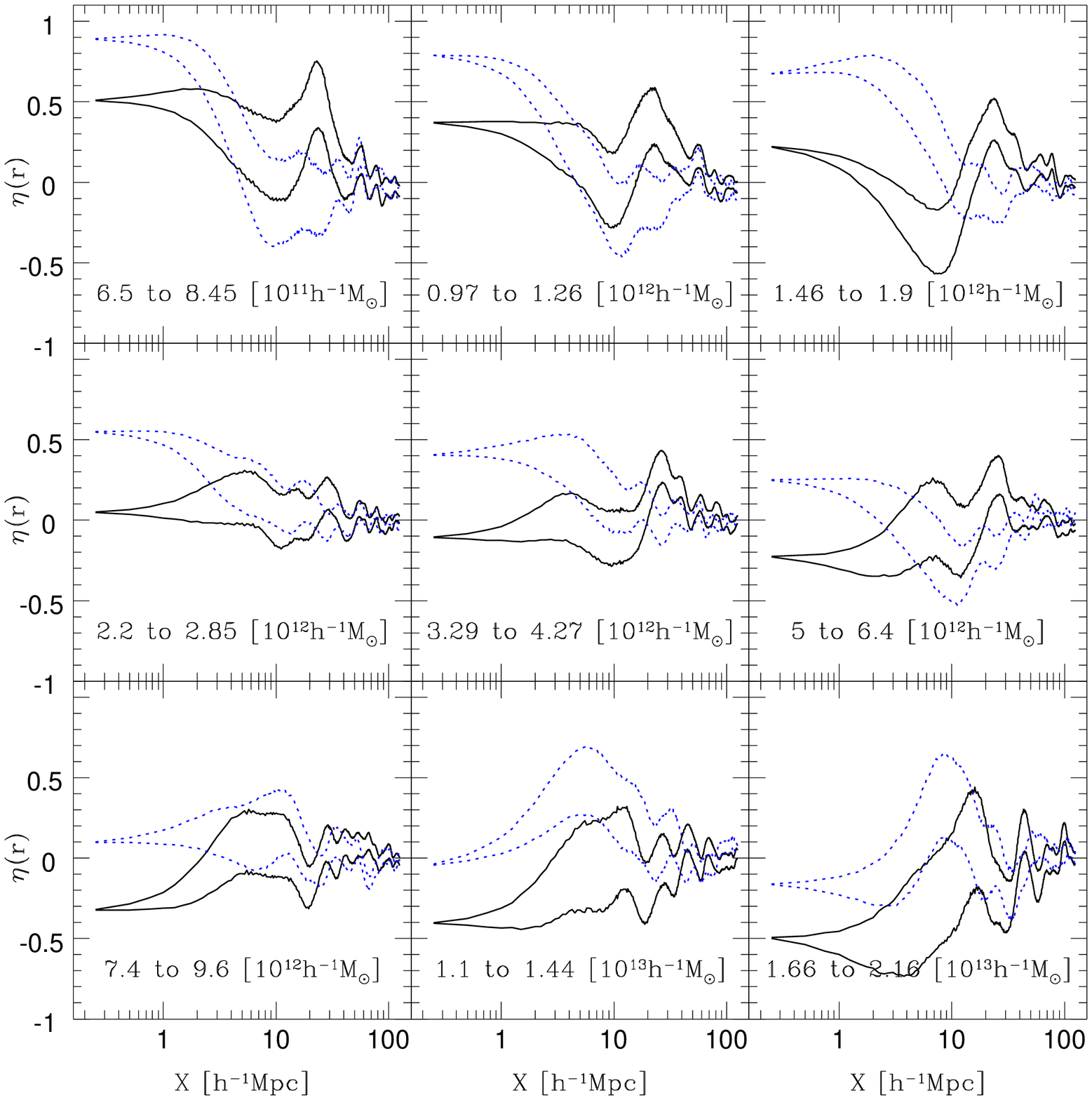, width=400pt, height=400pt}}
\caption{The mean dimensionality parameter, $\eta$, as a function of separation from 
particles making up  young (dotted blue lines representing $\pm\sigma$ deviations from the mean) and old haloes (solid black lines), all for $10\%$ oldest/youngest haloes.}
\label{fig:dimm}
\end{figure*}

Assembly bias may be caused by different environments of old and young halos.
We have experimented with cross correlating the bias with several statistical 
measure  of the environment. 
The most relevant measure  that we find is the ``dimensionality" of the 
density field in regions near halo particles at the initial time. This parameter 
is an indicator of the geometry of the structure developing 
at later times in those regions.
We show here that halo ages are strongly correlated with  the ``dimensionality" of initial fluctuation field as defined by  
\begin{equation}
\eta \equiv
\frac{\lambda_1+\lambda_2+\lambda_3}{\sqrt{\lambda_1^2+\lambda_2^2+\lambda_3^2}} 
\end{equation}
where $\lambda_i$ are the eigenvalues of 
the tensor $\partial \theta_l/\partial q_m$.
At the centers of   spherical, cylindrical and planar perturbations 
$\eta$ obtains the values 
 $\eta=\sqrt{3}$, $\sqrt{2}$ and $\eta=1$, respectively.
We have computed the mean value $\eta$ as a function of distance 
from particles making up young and old halos.  
The results are plotted in figure Fig.~\ref{fig:dimm}. 
Solid and dotted lines, respectively,  show  $\eta$ for old and young halos (two lines representing 
$\pm\sigma$ deviations from the mean, calculated as explained in \S3). 
This figure shows clearly that young haloes have an average higher dimensionality than old ones. 

\section{CONCLUSIONS}
\label{sec:sum}
We have shown that assembly bias of  halos  persists even in a simplified
description of gravitational dynamics like the  punctuated  Zel'dovich (PZ) approximation.
The PZ approximation  prevents the coasting away of particles in multi-streaming regions 
by coalescing objects that have come within a critical distance of 
each other. 
The PZ is fast,  simple to implement, and readily provides object merging trees.
This allows us to  study assembly bias in a large number of simulations (60 simulations, each of $512^3$ particles in a $(128h^{-1}{\rm Mpc})^3$ cubic box).   
The magnitude of the bias 
is  comparable to that found in full simulations. This implies that highly non-linear effects such as mass loss from halos in the vicinity of larger mass concentrations, may not be the dominant driver for 
assembly bias. 

We intend to apply the  PZ scheme to
the initial conditions used in the millennium simulation (Springel \etal 2005) and
compare the associated assembly  bias with the result of  Gao, Springel \& White (2005). This will yield a  better
quantitative assessment of  the role of highly non-linear effects. 

We have found a strong correlation between halo ages and the dimensionality 
of the nearby initial configuration. 
Young halos tend to form in regions of higher initial dimensionality 
than old halos. This is explained by the dependence of collapse time on 
dimensionality- a spherical perturbation collapses slower than a planar perturbation 
with the same initial density (Bertschinger \& Jain 1994).
Therefore,  assembly bias would be explained if one could show  
that regions of lower dimensionality are more correlated than those of higher dimensionality. This problem  could easily be tackled  numerically using realizations of random gaussian fields. 

\section*{ Acknowledgment }
This research is funded by the German-Israeli Foundation for Scientific Research
and Development and the Asher Space Research Fund.
We thank Vincent Desjacques for stimulating discussions
and Liron Gleser for help with the simulations and data analysis.
We are indebted to the referee, Ramin Skibba, for comments which 
helped improve the presentation of the paper.

\end{document}